\documentclass[12pt]{article}
\usepackage{amssymb}
\usepackage{amsfonts,amssymb,amsmath,amsthm}
\usepackage{epsfig}
\usepackage{slashed}
\usepackage{graphicx}
\usepackage{mathrsfs}
\usepackage{bbm}
\def\and{{\rm and}}

\begin{document}
\renewcommand{\thefootnote}{\fnsymbol{footnote}}
\begin{titlepage}
\begin{flushright}
USTC-ICTS-12-16
\end{flushright}

\vspace{10mm}
\begin{center}
{\Large\bf Investigation of non-Hermitian Hamiltonians in the Heisenberg picture}
\vspace{16mm}

{{\large Yan-Gang Miao${}^{1,2,}$\footnote{\em Corresponding author. E-mail: miaoyg@nankai.edu.cn}
and Zhen-Ming Xu${}^{1}$}\\

\vspace{6mm}
${}^{1}${\em School of Physics, Nankai University, Tianjin 300071, \\
People's Republic of China}

\vspace{3mm}
${}^{2}${\em Interdisciplinary Center for Theoretical Study,\\
University of Science and Technology of China, Hefei, Anhui 230026,\\
People's Republic of China}}

\end{center}

\vspace{10mm}
\centerline{{\bf{Abstract}}}
\vspace{6mm}
\noindent
The Heisenberg picture for non-Hermitian but $\eta$-pseudo-Hermitian Hamiltonian systems is suggested.
If a non-Hermitian but $\eta$-pseudo-Hermitian Hamiltonian leads to real second order equations of motion, though their first order Heisenberg equations of motion are complex, we can construct a Hermitian counterpart that gives the same second order equations of motion. In terms of a similarity transformation we verify the iso-spectral property of the Hermitian and non-Hermitian Hamiltonians and obtain the related eigenfunctions. This feature can be used to determine real eigenvalues for such non-Hermitian Hamiltonian systems. As an application, two new non-Hermitian Hamiltonians are constructed and investigated, where  
one is non-Hermitian and non-$PT$-symmetric and the other is non-Hermitian but $PT$-symmetric. 
Moreover, the complementarity and compatibility between our treatment and the $PT$ symmetry are discussed. 

\vskip 20pt
\noindent Number(s): 03.65.-w, 03.65.Ge, 03.65.Ta
\vskip 10pt
\noindent
\end{titlepage}

\newpage
\renewcommand{\thefootnote}{\arabic{footnote}}
\setcounter{footnote}{0}
\setcounter{page}{2}
\pagenumbering{arabic}

\section{Introduction}

It is well known that the operators of
physical observables are required to be (Dirac) Hermitian in order to have real eigenvalues in quantum mechanics.
However, the Hermiticity can be relaxed
to be $\eta$-pseudo Hermiticity or  $PT$ symmetry in non-Hermitian quantum mechanics, where $\eta$ is a linear Hermitian or an anti-linear anti-Hermitian
operator, and $P$ and $T$ stand for the parity and the time-reversal operators, respectively.
In the early 1940s, Dirac~\cite{Dirac} and Pauli~\cite{Pauli} introduced a non-Hermitian Hamiltonian and an $\eta$-dependent indefinite metric in the Hilbert space in order to deal with some divergence problems related to the unitarity of time evolution (conservation of probability).
Then, Lee and Wick~\cite{LW} applied in the late 1960s non-Hermitian Hamiltonians to quantum electrodynamics for keeping unitarity of the $S$-matrix. Later, many other authors revealed~\cite{BFMC} in different areas of research that a non-Hermitian Hamiltonian could have real eigenvalues under specific conditions. Till 1998 Bender and Boettcher~\cite{BBR} constructed a special class of non-Hermitian Hamiltonians and studied the relationship between the $PT$ symmetry and real eigenvalues. Since then, a large amount of work has been done, mainly on the $PT$-symmetric Quantum Mechanics~\cite{Bender} and $\eta$-pseudo-Hermitian Quantum Mechanics~\cite{Ali}.

Normally, a non-Hermitian Hamiltonian is analyzed in terms of Schr\"odinger equations in the $\eta$-pseudo-Hermitian and the $PT$-symmetric
quantum theories. Because of the non-Hermiticity of Hamiltonians, new concepts are introduced, such as indefinite and positive definite metrics, biorthonormal bases, and modified inner products, etc.
Such an analysis has been demonstrated in detail in the review articles~\cite{Bender,Ali} for various non-Hermitian quantum systems.

In this paper, instead of solving Schr\"odinger equations, we analyze eigenvalues of non-Hermitian quantum systems from the point of view of Heisenberg equations of motion. To this end, the Heisenberg picture for non-Hermitian but $\eta$-pseudo-Hermitian Hamiltonians is given as a basis. Then, the first order Heisenberg equations of motion can be derived and they are found to be complex in general. If the corresponding second  order equations of motion are real,
we deduce such a Hermitian Hamiltonian that leads to the same second  order equations of motion.
Through a non-unitary similarity transformation we verify that the Hermitian and non-Hermitian Hamiltonians have the same spectrum. In this way, we can determine that the non-Hermitian Hamiltonian has real eigenvalues. 
As an application, two new non-Hermitian Hamiltonians are constructed and investigated, where  
one is non-Hermitian and non-$PT$-symmetric and the other is non-Hermitian but $PT$-symmetric.
We apply the Heisenberg picture to a non-$PT$-symmetric model and thus extend the investigation of real eigenvalues of non-Hermitian systems from $PT$-symmetric Hamiltonians to non-$PT$-symmetric ones.  As to our second model that is $PT$-symmetric, we wish to show that our treatment is also valid to a $PT$-symmetric system.

This paper is arranged as follows.
In the following section, the Heisenberg picture is provided for non-Hermitian but $\eta$-pseudo-Hermitian Hamiltonians. We find that the usual formulation of Heisenberg equations of motion maintains for the non-Hermitian case if a suitably modified inner product is introduced.
Two new non-Hermitian Hamiltonians are proposed and analyzed in sections 3 and 4, respectively. Although their first order Heisenberg equations of motion are highly non-trivially complex, the corresponding second order equations of motion are shown to be real. Then, the  iso-spectral Hermitian Hamiltonians that give the same second order equations of motion are deduced,  and the existence of real eigenvalues for the two non-Hermitian Hamiltonians can thus be determined.   
Finally, section 5 is devoted to a conclusion, where the complementarity and compatibility between our treatment and the $PT$ symmetry are briefly discussed and a possible application of our models in optics is envisioned.

\section{Heisenberg picture for non-Hermitian Hamiltonians}
For a non-Hermitian, $H^{\dag} \neq H$,  but $\eta$-pseudo-Hermitian Hamiltonian, 
\begin{equation}
H^{\ddagger} \equiv {\eta}^{-1}H^{\dag}{\eta} =H, \label{pseudoHermi}
\end{equation}
where $H$ does not  depend explicitly on time and $\eta$,  a metric named by Pauli, is a linear Hermitian operator, the modified inner product is defined~\cite{Pauli} as 
\begin{equation}
\langle \psi(t) | \psi(t) \rangle_{\eta} \equiv \langle \psi(t)| {\eta} | \psi(t) \rangle,\label{MIP}
\end{equation}
where the wave function at any time $\psi(t)$ satisfies 
the Schr\"odinger equation, $i\hbar\frac{\partial}{\partial t}\psi(t)=H \psi(t)$, and it can be expressed by  the wave function at the initial time as follows:
\begin{equation}
\psi(t)= \exp(-iHt/\hbar)\psi(0).\label{psit}
\end{equation}
For a physical observable $O$, in accordance with the modified inner product eq.~(\ref{MIP}), its average value now takes the following form in the Schr\"odinger picture,
\begin{equation}
\langle O \rangle_{\rm Av}^{\rm S}=\langle \psi(t) | O | \psi(t) \rangle_{\eta}=\langle \psi(t) | {\eta}\,O | \psi(t) \rangle,\label{Oav}
\end{equation}
which is definitely real because $O$, as a physical observable, has the same $\eta$-pseudo Hermiticity as $H$,  $O={\eta}^{-1}O^{\dag}{\eta}$.
Substituting eq.~(\ref{psit}) into eq.~(\ref{Oav}) and using eq.~(\ref{pseudoHermi}), we rewrite the average value of the operator $O$ to be
\begin{eqnarray}
\langle O \rangle_{\rm Av}^{\rm S} &=& \langle \psi(0) | \exp(+iH^{\dagger}t/\hbar)\,\eta\,O\exp(-iHt/\hbar) | \psi(0) \rangle \nonumber \\
&=& \langle \psi(0) | {\eta}\left\{{\eta}^{-1}\exp(+iH^{\dagger}t/\hbar)\,\eta\right\}O\exp(-iHt/\hbar) | \psi(0) \rangle \nonumber \\
&=& \langle \psi(0) | \exp(iHt/\hbar)\,O\exp(-iHt/\hbar) | \psi(0) \rangle_{\eta}.\label{OavS}
\end{eqnarray}

Alternatively, one can introduce the Heisenberg picture where the average value of the time dependent physical observable  $O(t)$ has the form, 
\begin{equation}
\langle O(t) \rangle_{\rm Av}^{\rm H} = \langle \psi(0) | O(t) | \psi(0) \rangle_{\eta}.\label{OavH}
\end{equation}
In accordance with the principle that the average value of an arbitrary physical observable is independent of the choice of pictures, that is, 
\begin{equation}
\langle O \rangle_{\rm Av}^{\rm S}=\langle O(t) \rangle_{\rm Av}^{\rm H},\label{SequalH}
\end{equation}
we obtain the time dependent observable in terms of the time independent one by comparing eq.~(\ref{OavS}) and eq.(\ref{OavH}),
\begin{equation}
O(t) = \exp(iHt/\hbar)O\exp(-iHt/\hbar) \label{Ot}.
\end{equation}
The time evolution of $O(t)$, {\em i.e.}, the Heisenberg equation of motion thus has the usual formulation,
\begin{equation}
\dot{O}(t) =\frac{1}{i\hbar}[O(t), H].\label{Heisenbergequ}
\end{equation}
Consequently, eqs.~(\ref{OavH}) and (\ref{Heisenbergequ}) give the Heisenberg picture under the definition of modified inner products (eq.~(\ref{MIP})). 
Note that here the time dependent operator  (sometimes called Heisenberg operator) $O(t)$ has the $\eta$-pseudo Hermiticity, 
\begin{equation}
O(t)={\eta}^{-1}\{O(t)\}^{\dag}{\eta},\label{OtpseudoH}
\end{equation} 
due to the $\eta$-pseudo Hermiticity of time independent operators, $H={\eta}^{-1}H^{\dag}{\eta}$ and $O={\eta}^{-1}O^{\dag}{\eta}$.  Further, the relationship of commutators between the two pictures can be derived,
\begin{equation}
[{O}_1(t), O_2(t)] =\exp(iHt/\hbar)[O_1, O_2]\exp(-iHt/\hbar).\label{OtOcommu}
\end{equation}
This leads to the maintenance of Heisenberg commutation relations for canonical time dependent operators. For instance, to the coordinate and the momentum, $x(t)$ and $p(t)$ satisfy 
\begin{equation}
[x(t), p(t)]=i\hbar, \qquad [x(t), x(t)]=0=[p(t), p(t)], \label{xtptcomm}
\end{equation}
if the time independent counterparts obey the (canonical) Heisenberg commutation relations.

\section{The non-Hermitian and non-$PT$-symmetric model}

In general, the real equations of motion are closely related with the Hermiticity of Hamiltonians, which provides the basis of our analysis. For a quantum dynamical system described by a Hermitian Hamiltonian,\footnote{Time is hidden for Heisenberg operators in the following context for the sake of convenience.} $H_{\rm Hermitian}=\frac{p^2}{2m} + V(x)$,  where $V(x)$ is an arbitrary  real potential that is usually required to be differentiable with respect to $x$, the Hamiltonian leads of course  to a real second order equation of motion, $m\ddot{x} + V^{\prime}(x)=0$. On the contrary, one can deduce a Hermitian Hamiltonian from a real second order equation of motion and fix the Hamiltonian up to a real constant. Consequently, if a non-Hermitian Hamiltonian gives a real second order equation of motion, its Hermitian counterpart can be deduced from the real second order equation of motion. 
The iso-spectrum of the non-Hermitian Hamiltonian and its Hermitian counterpart can be verified through a similarity transformation. In this way, one determines the existence of real eigenvalues for  such a non-Hermitian Hamiltonian.

Now we turn to our first model. Let $\sum_{k=0}^{\infty}c_k x^{k+n}$ be a general series, where the index $n$ can take zero or any of positive integers, and $c_k$'s related to this index are real parameters. The radius of convergence is defined as $R\equiv \lim_{k \to \infty} \left | \frac{c_k}{c_{k+1}}\right |$, and the range of $x$'s average values is required to be less than $R$. Note that $R$ can take infinity, for example, when this series is the Bessel function of the first kind of order $n$, $J_n(x)$, where $n=0, 1, 2, \cdots$. This means that this series can cover special functions. By using the series, we construct the following non-Hermitian Hamiltonian,
\begin{equation}
H=\frac{p^2}{2m} + V(x) + \frac{i}{2}\left\{ \left(\sum_{k=0}^{\infty} {c_k}x^{k+n}\right)p + p \left(\sum_{k=0}^{\infty} {c_k}x^{k+n}\right)\right\}, \label{generalH}
\end{equation}
where  $(x, p)$ is a pair
of canonical coordinate and momentum that satisfies the Heisenberg commutation relations given by eq.~(\ref{xtptcomm}). Note that this non-Hermitian Hamiltonian is not $PT$ symmetric in general, which can be seen  on the one hand because $V(x)$ is  generally not an even function, {\em i.e.}, $(PT)^{-1}V(x)(PT)=V(-x)\neq V(x)$, and on the other hand because the series usually contains powers of even numbers.

According to the Heisenberg picture established in the above section, we  derive the Heisenberg equations of motion for the non-Hermitian and non-$PT$-symmetric Hamiltonian in light of eqs.~(\ref{Heisenbergequ}), (\ref{xtptcomm}), and (\ref{generalH}),
\begin{eqnarray}
\dot{x}&=&\frac{p}{m} +  i \sum_{k=0}^{\infty}{c_k}x^{k+n},  \label{GHeisenbergEq1} \\ 
\dot{p}&=&-V^{\prime}(x)- \frac{i}{2}\left\{ \left(\sum_{k=0}^{\infty} {c_k}x^{k+n}\right)^{\prime}p + p \left(\sum_{k=0}^{\infty} {c_k}x^{k+n}\right)^{\prime}\right\},\label{GHeisenbergEq2}
\end{eqnarray}
where the prime stands for the derivative with respect to  $x$.
Although the Heisenberg equations of motion are highly non-trivially complex, surprisingly, we find\footnote{See Appendix A for the detailed derivation.} that the quantum second order equation of motion for the coordinate $x$ is  real when  eliminating the momentum $p$,
\begin{equation}
m\ddot{x} + V^{\prime}(x) +\frac{m}{2}\left\{\left(\sum_{k=0}^{\infty}{c_k}x^{k+n}\right)^2\right\}^{\prime}=0. \label{GEOM}
\end{equation}
Correspondingly, we can deduce a Hermitian Hamiltonian that gives the same real equation of motion,
\begin{equation}
h=\frac{p^2}{2m} + V(x)  +\frac{m}{2}\left(\sum_{k=0}^{\infty}{c_k}x^{k+n}\right)^2, \label{GHH}
\end{equation}
which can be fixed up to a real constant that has been set be zero.

Now we verify that the non-Hermitian and non-$PT$-symmetric Hamiltonian (eq.~(\ref{generalH})) can be converted into its Hermitian (iso-spectral) counterpart (eq.~(\ref{GHH})) by a non-unitary similarity transformation. Set 
\begin{eqnarray}
H\Phi=E\Phi, \qquad h\phi=E\phi, \label{Hheigen}
\end{eqnarray}
{\em i.e.}, $H$ and $h$ have the same eigenvalues $E$, and $\Phi$ and $\phi$ are their eigenfunctions, respectively.
By using the Baker-Campbell-Hausdorff formula we can find out such a non-unitary operator $\Omega$,
\begin{equation}
\Omega = \exp \left(-\frac{m}{\hbar}\sum_{k=0}^{\infty}\frac{c_k}{k+n+1}x^{k+n+1}\right), \label{Otrans}
\end{equation}
which is a linear operator, that it connects the two Hamiltonians as follows:
\begin{eqnarray}
h=\Omega H {\Omega}^{-1}. \label{hH} 
\end{eqnarray}
The detailed derivation is given in Appendix B. Further considering eq.~(\ref{Hheigen}), we get the relation between the two sets of eigenfunctions  as follows:
\begin{eqnarray}
\phi &=& \Omega \Phi. \label{connef}
\end{eqnarray}

Consequently, we determine the existence of real eigenvalues for the non-Hermitian and non-$PT$-symmetric Hamiltonian (eq.~(\ref{generalH})) in the Heisenberg picture when all $c_k$ are real parameters. 
As to the imaginary interacting potential in this Hamiltonian, i.e. the third term, 
its effect, when we focus only on eigenvalues, is equivalent to the contribution of the real potential $\frac{m}{2}\left(\sum_{k=0}^{\infty}{c_k}x^{k+n}\right)^2$, see eq.~(\ref{GHH}).

Before ending this section,  
we emphasize that the non-Hermitian Hamitonian eq.~(\ref{generalH}) inherently possesses the $\eta$-pseudo-Hermiticity, which is the prerequisite for us to adopt the Heisenberg picture. Taking into account  eq.~(\ref{hH}) and the Hermiticity of $h$, we can verify the $\eta$-pseudo-Hermiticity of $H$,
\begin{equation}
H^{\ddagger}\equiv {\eta}^{-1}H^{\dagger}\eta = H, \label{pseudoHrelation}
\end{equation}
where $\eta$ is expressed in terms of  $\Omega$ as follows:
\begin{equation}
\eta = {\Omega}^{\dagger}\Omega. \label{etaOmega}
\end{equation}
Note that the metric operator ${\eta}$ is both Hermitian and ${\eta}$-pseudo-Hermitian self-adjoint,  that is, it is obvious to see ${\eta}^{\dagger}={\eta}$, and ${\eta}^{\ddagger}\equiv {\eta}^{-1}{\eta}^{\dagger}\eta = {\eta}$.

\section{The non-Hermitian but $PT$-symmetric model}
Let us turn to the second model that is  
non-Hermitian but $PT$-symmetric. Its Hamiltonian is given as follows:
\begin{equation}
{\tilde H}=\frac{A}{2}x^2 + V(p) + \frac{i}{2}\left\{ \left(\sum_{k=0}^{\infty} {a_k}p^{k+n}\right)x + x \left(\sum_{k=0}^{\infty} {a_k}p^{k+n}\right)\right\}, \label{generalH2}
\end{equation}
where $A$ is a positive constant with the dimension of $[M][T]^{-2}$, $V(p)$ is an arbitrary  real potential of momentum that is usually required to be differentiable to $p$, $\sum_{k=0}^{\infty} {a_k}p^{k+n}$ is a  general series of momentum $p$, where ${a_k}$'s are real constants,  the index $n$ takes zero or any of positive integers, and $(x, p)$ is a pair
of canonical coordinate and momentum that satisfies the Heisenberg commutation relations eq.~(\ref{xtptcomm}). This Hamiltonian may be understood in the momentum representation of the Hamiltonian (eq.~(\ref{generalH})) constructed in the above section, and hence its $PT$ symmetry is restored due to the exchange between coordinate and momentum in eq.~(\ref{generalH}).

Following the same procedure as in the above section, {\em i.e.}, using the Heisenberg picture, we at first derive the Heisenberg equations of motion for the coordinate and the momentum from the non-Hermitian $PT$-symmetric Hamiltonian eq.~(\ref{generalH2}), 
\begin{eqnarray}
\dot{x}&=&V^{\prime}(p)+\frac{i}{2}\left\{ \left(\sum_{k=0}^{\infty} {a_k}p^{k+n}\right)^{\prime}x + x \left(\sum_{k=0}^{\infty} {a_k}p^{k+n}\right)^{\prime}\right\},  \label{tildeGHeisenbergEq1} \\ 
\dot{p}&=&-Ax -  i \sum_{k=0}^{\infty}{a_k}p^{k+n},\label{tildeGHeisenbergEq2}
\end{eqnarray}
where the prime stands here for the derivative with respect to the momentum $p$. 
Similar to eqs.~(\ref{GHeisenbergEq1}) and (\ref{GHeisenbergEq2}), eqs.~(\ref{tildeGHeisenbergEq1}) and (\ref{tildeGHeisenbergEq2}) are also highly non-trivially complex. However, by eliminating the coordinate we then obtain the real second order equation of motion with respect to the momentum $p$,
\begin{equation}
\frac{1}{A}\ddot{p} + V^{\prime}(p) +\frac{1}{2A}\left\{\left(\sum_{k=0}^{\infty}{a_k}p^{k+n}\right)^2\right\}^{\prime}=0. \label{EOM2}
\end{equation}
Using this real equation of motion we deduce the corresponding Hermitian Hamiltonian,
\begin{equation}
{\tilde h}=\frac{A}{2}x^2 + V(p) + \frac{1}{2A}\left(\sum_{k=0}^{\infty}{a_k}p^{k+n}\right)^2. \label{GHH2}
\end{equation}
We point out that ${\tilde h}$ has real eigenvalues that are equal to the eigenvalues of ${\tilde H}$. That is, the property of iso-spectrum 
can be verified by the following relations, 
\begin{eqnarray}
& &{\tilde h}={\tilde{\Omega}} \tilde H {\tilde{\Omega}}^{-1}, \\
& &\tilde \phi = {\tilde{\Omega}} \tilde \Phi, \label{tildeHh2}
\end{eqnarray}
where the operator ${\tilde{\Omega}}$ is found in light of the Baker-Campbell-Hausdorff formula to be,
\begin{equation}
{\tilde{\Omega}} = \exp \left(\frac{1}{A\hbar}\sum_{k=0}^{\infty}\frac{a_k}{k+n+1}p^{k+n+1}\right),\label{tildeO}
\end{equation}
and $\tilde H \tilde \Phi=\tilde E \tilde \Phi$ and $\tilde h \tilde \phi=\tilde E \tilde \phi$ are set like eq.~(\ref{Hheigen}). As a result, we determine the existence of real eigenvalues for the non-Hermitian $PT$-symmetric Hamiltonians (eq.~(\ref{generalH2})) in terms of the Heisenberg picture. 
Similar to the model in the above section, to the imaginary interacting potential in eq.~(\ref{generalH2}), $\frac{i}{2}\left\{ \left(\sum_{k=0}^{\infty} {a_k}p^{k+n}\right)x + x \left(\sum_{k=0}^{\infty} {a_k}p^{k+n}\right)\right\}$, its effect equals the contribution of the real potential $\frac{1}{2A}\left(\sum_{k=0}^{\infty}{a_k}p^{k+n}\right)^2$ in the aspect of eigenvalues, see eq.~(\ref{GHH2}).

Now it is necessary to 
emphasize that the non-Hermitian Hamitonian eq.~(\ref{generalH2}) inherently has the $\tilde{\eta}$-pseudo-Hermiticity, which is the prerequisite for us to adopt the Heisenberg picture. Considering  eq.~(\ref{tildeHh2}) and the Hermiticity of $\tilde h$, we can verify the $\tilde{\eta}$-pseudo-Hermiticity of $\tilde H$,
\begin{equation}
{\tilde H}^{\ddagger}\equiv {\tilde{\eta}}^{-1}{\tilde H}^{\dagger}\tilde{\eta} = \tilde H, \label{tildepseudoHrelation}
\end{equation}
where $\tilde{\eta}$ is expressed in terms of  $\tilde \Omega$ as follows:
\begin{equation}
\tilde{\eta} = {\tilde \Omega}^{\dagger}\tilde \Omega. \label{tildeetaOmega}
\end{equation}
This metric operator ${\tilde{\eta}}$ is both Hermitian and ${\tilde{\eta}}$-pseudo-Hermitian self-adjoint,  that is, it is easy to verify  ${\tilde{\eta}}^{\dagger}={\tilde{\eta}}$, and ${\tilde{\eta}}^{\ddagger}\equiv {\tilde{\eta}}^{-1}{\tilde{\eta}}^{\dagger}\tilde{\eta} = {\tilde{\eta}}$.

In addition,  one can deduce the existence of real eigenvalues for $\tilde H$ (eq.~(\ref{generalH2})) in accordance with the non-Hermitian $PT$-symmetric quantum theory~\cite{Bender}, which shows that our analysis is compatible with the $PT$ symmetry.

Before ending this section, we mention that the two known models, the Swanson model~\cite{Swanson} and the Pais-Uhlenbeck oscillator model~\cite{PU}, can also be dealt with in the Heisenberg picture.

For the Swanson model, it is just the special case of eq.~(\ref{generalH}) with $V(x)$ the potential of the harmonic oscillator, $n=0$, and the only non-vanishing coefficient $c_1$.

For the Pais-Uhlenbeck oscillator model, it has a little difficulty to treat because its equation of motion is fourth order that gives rise to the appearance of negative norms. However, this problem has been overcome from the $PT$-symmetric point of view~\cite{Bender1}. Alternatively, it can also be solved from the $PT$-pseudo-Hermitian point of view~\cite{LM}. Here we briefly revisit this model in the Heisenberg picture by starting with its $PT$-pseudo-Hermitian Hamiltonian,
\begin{equation}
H_{\rm PU}=\frac{p^2_1}{2m} + \frac{1}{2}ma^2_1 x^2_1 + \frac{p^2_2}{2m}  + \frac{1}{2}ma^2_2 x^2_2 + i\frac{a_3}{2ma_1 a_2}p_1p_2,\label{PU2}
\end{equation}
where $m$ is the mass of an anisotropic two-dimensional oscillator, and $a_1$, $a_2$, and $a_3$ are non-vanishing real constants with
the inequality, $|a_3| < |a^2_1-a^2_2|$. This model  was constructed by adding an imaginary interacting term proportional to $ip_1p_2$ to the Hamiltonian of a free anisotropic planar oscillator, for the details, see ref.~\cite{LM}. We can derive its Heisenberg equations of motion for both the coordinate operators and the momentum operators, which are coupled complex equations. When eliminating the momenta and decoupling the two spatial dimensions, we arrive at the same real fourth order equation of motion for both $x_1$ and $x_2$. Exactly following the procedure utilized in our two models (eqs.~(\ref{generalH}) and (\ref{generalH2})), we find that $H_{\rm PU}$ can be converted into its Hermitian counterpart $h_{\rm PU}$ and both of them have the same eigenvalues through the following transformations,
\begin{eqnarray}
& &{h}_{\rm PU}={{\Omega}_{\rm PU}}  H_{\rm PU} {\Omega}_{\rm PU}^{-1}, \\
& &  \phi_{\rm PU} = {\Omega}_{\rm PU}\Phi_{\rm PU}, \label{PUHh2}
\end{eqnarray}
where we have set $ H_{\rm PU}  \Phi_{\rm PU}= E_{\rm PU}  \Phi_{\rm PU}$ and $ h_{\rm PU}  \phi_{\rm PU}= E_{\rm PU}  \phi_{\rm PU}$. The relevant results are listed as follows:
\begin{eqnarray}
& & \Omega_{\rm PU} = \exp \left(-\frac{\alpha x_1 p_2-\beta x_2 p_1}{2\hbar}\right),  \\
& & h_{\rm PU}=\frac{p^2_1}{2m_1} + \frac{1}{2}m_1 \omega^2_1 x^2_1 + \frac{p^2_2}{2m_2}  + \frac{1}{2}m_2 \omega^2_2 x^2_2,
\end{eqnarray}
where the newly introduced parameters, if $a_1 > a_2$, are defined as
\begin{eqnarray}
& &\alpha=\frac{a_1}{a_2}\ln\left(\frac{a_3+|a^2_1-a^2_2|}{\sqrt{(a^2_1-a^2_2)^2-a^2_3}}\right), \qquad \beta=\frac{a_2}{a_1}\ln\left(\frac{a_3+|a^2_1-a^2_2|}{\sqrt{(a^2_1-a^2_2)^2-a^2_3}}\right), \nonumber \\
& & m_1=\frac{m}{1-\frac{U}{2a^2_1}},  \qquad m_2=\frac{m}{1+\frac{U}{2a^2_2}}, \nonumber \\
& & \omega_1^2=a^2_1\left(1-\frac{U}{2a^2_1}\right),  \qquad \omega_2^2=a^2_2\left(1+\frac{U}{2a^2_2}\right), \nonumber \\
& & U=\frac{a^2_3+(a^2_2-a^2_1)\left(|a^2_1-a^2_2|-\sqrt{(a^2_1-a^2_2)^2-a^2_3}\right)}{\sqrt{(a^2_1-a^2_2)^2-a^2_3}}.
\end{eqnarray}
Note that two masses ($m_1$ and $m_2$) and two angular frequency squares ($\omega^2_1$ and $\omega^2_2$) are positive. If $a_1 < a_2$, the corresponding  parameters can be obtained by the permutation between $a_1$ and $a_2$ in the above seven parameters, and the new masses and angular frequency squares are still positive.

\section{Summary}

In this paper we propose a suggestion --- investigation of non-Hermitian Hamiltonians in the Heisenberg picture --- to determine the existence of real eigenvalues for a non-Hermitian Hamiltonian system. The analysis can be fulfilled in the Heisenberg picture we establish for a non-Hermitian but $\eta$-pseudo-Hermitian Hamiltonian. We apply this proposal to two complicated non-Hermitian Hamiltonians, where one is non-Hermitian non-$PT$-symmetric, and the other is non-Hermitian $PT$-symmetric. The models are shown to have real second order equations of motion, and their Hermitian counterparts of Hamiltonians are deduced. 
The iso-spectrum is verified by a non-unitary similarity transformation between a non-Hermitian Hamiltonian and a Hermitian one.

Our analysis is complementary to the $PT$-symmetric quantum mechanics. 
That is,  to a model that has real equations of motion but no $PT$ symmetry, such as eq.~(\ref{generalH}), one can analyze it as stated in the present paper. On the other hand, to a model that has $PT$ symmetry but no real equations of motion, such as the model with the imaginary cubic potential $ix^3$, one can follow the non-Hermitian $PT$-symmetric quantum theory~\cite{Bender}. Moreover, 
our discussion about eq.~(\ref{generalH2}) shows that our proposal is also compatible with the $PT$ symmetry through investigating the non-Hermitian Hamiltonian systems that possess both the real equations of motion and the $PT$ symmetry.  

At last, we envision a possible application of our models in optics.

First of all, let us give a brief review of an interesting optical model that is related to non-Hermitian potentials. Based on the formal equivalence between the Schr\"odinger equation in quantum mechanics and the wave equation in optics, a $PT$-symmetric periodic potential was proposed~\cite{Makris} within the framework of optics,
\begin{eqnarray}
V_{\rm optics}(x) \propto \cos^2x + i V_0\sin (2x), \label{opticalV}
\end{eqnarray}
where $V_0$ is a real constant. This complex potential describes a one-dimensional planar inhomogeneous configuration with a complex refractive index distribution, where the real part stands for the real index profile of the lattice structure and the imaginary part represents the gain or loss component.
Beam dynamics was examined in terms of this complex potential and some intriguing characteristics were revealed in double refraction, power oscillations, eigenfunction unfolding, and nonreciprocal diffraction patterns.

Now we turn to our model given by eq.~(\ref{generalH}). When we take a special case, i.e. $V(x) \propto \cos^2x$, $c_k \propto V_0\frac{(-1)^k{2}^{n+k}}{(2k+1)!}$, and $n=k+1$, our complex potential takes a simpler form  as follows: 
\begin{eqnarray}
V_{\rm our \, \,model}(x) \propto \cos^2x + i \frac{V_0}{2}\left\{\sin(2x)p + p \sin(2x)\right\}, \label{ourV}
\end{eqnarray}
whose real part is same as but whose imaginary part is different from that of eq.~(\ref{opticalV}).
This potential describes, in the sense of the formal equivalence between the quantum mechanical equation and the optical wave equation, a different lattice structure from that depicted by eq.~(\ref{opticalV}), in particular, its imaginary refractive index depends on the interaction between coordinates and momenta, while no such an interaction exists in the imaginary part of eq.~(\ref{opticalV}). On the other hand, our complex potential is non-$PT$-symmetric, but $\eta$-pseudo Hermitian (here $\eta$ is a special case of eq.~(\ref{etaOmega})), with which one may extend the study of beam dynamics from a $PT$-symmetric system to a non-$PT$-symmetric one. When such a complex potential is applied to beam dynamics, we suppose that it may probably lead to rich characteristics beyond those that have been pointed out by ref.~\cite{Makris}. 

\section*{Acknowledgments}
Y-GM would like to thank J.-X. Lu for warm hospitality when he visited the Interdisciplinary Center for Theoretical Study (ICTS),
University of Science and Technology of China (USTC). 
This work was supported in part by the National Natural Science Foundation of China under grant No.11175090 and 
by the Ministry of Education of China under grant No.20120031110027.
At last, the authors would like to thank the anonymous referees and the editor for their helpful comments that indeed greatly improve this work.

\section*{Appendix A \hspace{.24cm}Derivation of eq.~(\ref{GEOM})}
Making derivative to eq.~(\ref{GHeisenbergEq1}) with respect to time, we have
\begin{equation}
\ddot{x}=\frac{\dot{p}}{m}+i  \sum_{k=0}^{\infty}{c_k}\left(\dot{x}x^{k+n-1}+x\dot{x}x^{k+n-2}+ \cdots + x^{k+n-2}\dot{x}x+x^{k+n-1}\dot{x}\right).\label{eq1}
\end{equation}
Substituting eq.~(\ref{GHeisenbergEq2}) into eq.~(\ref{eq1}),  we obtain
\begin{eqnarray}
\ddot{x} &=& -\frac{1}{m} V^{\prime}(x)- \frac{i}{2m}\left\{\left(\sum_{k=0}^{\infty} {c_k}x^{k+n}\right)^{\prime}p + p \left(\sum_{k=0}^{\infty} {c_k}x^{k+n}\right)^{\prime}\right\} \nonumber \\
& & + i  \sum_{k=0}^{\infty}{c_k}\left(\dot{x}x^{k+n-1}+x\dot{x}x^{k+n-2}+ \cdots + x^{k+n-2}\dot{x}x+x^{k+n-1}\dot{x}\right). \label{eq2}
\end{eqnarray}
Next, substituting eq.~(\ref{GHeisenbergEq1}) into the third term of eq.~(\ref{eq2}) and considering the Heisenberg commutation relations (eq.~(\ref{xtptcomm})),  we reduce this term to be
\begin{equation}
\frac{i}{2m}\left\{\left(\sum_{k=0}^{\infty} {c_k}x^{k+n}\right)^{\prime}p + p \left(\sum_{k=0}^{\infty} {c_k}x^{k+n}\right)^{\prime}\right\}-\frac{1}{2}\left\{\left(\sum_{k=0}^{\infty}{c_k}x^{k+n}\right)^2\right\}^{\prime}.\label{eq3}
\end{equation}
Combining  eq.~(\ref{eq2}) with eq.~(\ref{eq3}), we see the cancellation of the imaginary terms and thus derive the real second order equation of motion --- eq.~(\ref{GEOM}).

\section*{Appendix B \hspace{.24cm}Verification of eq.~(\ref{hH})}

Using eq.~(\ref{Otrans}) and eq.~(\ref{xtptcomm}), we make the similarity transformation to  the three terms in eq.~(\ref{generalH}),
\begin{eqnarray}
& &\Omega \left\{\frac{p^2}{2m}\right\} {\Omega}^{-1} = \frac{p^2}{2m} - \frac{i}{2}\left\{ \left(\sum_{k=0}^{\infty} {c_k}x^{k+n}\right)p + p \left(\sum_{k=0}^{\infty} {c_k}x^{k+n}\right)\right\} - \frac{m}{2}\left(\sum_{k=0}^{\infty}{c_k}x^{k+n}\right)^2,
\nonumber \\
& & \Omega \left\{V(x)\right\} {\Omega}^{-1} =  V(x),  \nonumber \\
& & \Omega \left\{\frac{i}{2}\left\{ \left(\sum_{k=0}^{\infty} {c_k}x^{k+n}\right)p + p \left(\sum_{k=0}^{\infty} {c_k}x^{k+n}\right)\right\}\right\} {\Omega}^{-1}  \nonumber \\
& & = \frac{i}{2}\left\{ \left(\sum_{k=0}^{\infty} {c_k}x^{k+n}\right)p  + p \left(\sum_{k=0}^{\infty} {c_k}x^{k+n}\right)\right\} + m\left(\sum_{k=0}^{\infty}{c_k}x^{k+n}\right)^2.\nonumber
\end{eqnarray}
By adding the terms in the above three equations in the manner of left to left and right to right, we obtain
\begin{eqnarray}
\Omega H {\Omega}^{-1} = h,
\end{eqnarray}
where $h$ is the Hermitian Hamiltonian given in eq.~(\ref{GHH}).



\begin{thebibliography}{99}

\bibitem{Dirac}P.A.M. Dirac, {\em The physical interpretation of quantum mechanics}, Proc. Roy. Soc. Lond. {\bf A 180} (1942) 1.

\bibitem{Pauli}W. Pauli, {\em On Dirac's new method of field quantization}, Rev. Mod. Phys. {\bf 15} (1943) 175.

\bibitem{LW}T.D. Lee and G.C. Wick, {\em Negative metric and the unitarity of the S-matrix}, Nucl. Phys. {\bf B 9} (1969) 209.

\bibitem{BFMC}R. Brower, M. Furman, and M. Moshe, {\em Critical exponents for the Reggeon quantum spin model},
Phys. Lett. {\bf B 76} (1978) 213;\\
B.C. Harms, S.T. Jones, and C.-I Tan, {\em Complex energy spectra in Reggeon quantum mechanics with quartic interactions}, Nucl. Phys. {\bf B 171} (1980) 392; {\em New structure in the energy spectrum of Reggeon quantum mechanics with quartic couplings}, Phys.
Lett. {\bf B 91} (1980) 291;\\
E. Caliceti, S. Graffi, and M. Maioli, {\em Perturbation theory of odd anharmonic oscillators}, Comm. Math. Phys. {\bf 75} (1980) 51;\\
A.A. Andrianov, {\em The large N expansion as a local perturbation theory}, Ann. Phys. (N.Y.) {\bf 140} (1982) 82;\\
T. Hollowood, {\em Solitons in affine Toda field theories}, Nucl. Phys. {\bf B 384} (1992) 523 [arXiv:hep-th/9110010];\\
F.G. Scholtz, H.B. Geyer, and F.J.W. Hahne, {\em Quasi-Hermitian operators in quantum mechanics and the variational principle},
Ann. Phys. (N.Y.) {\bf 213} (1992) 74.

\bibitem{BBR}C.M. Bender and S. Boettcher, {\em Real spectra in non-Hermitian Hamiltonians having PT symmetry},
Phys. Rev. Lett. {\bf 80} (1998) 5243 [arXiv:math-ph/9712001].

\bibitem{Bender}See, for example, the review article and the references therein.\\
 C.M. Bender, {\em Making sense of non-Hermitian Hamiltonians}, Rep. Prog. Phys. {\bf 70} (2007) 947 [arXiv:hep-th/0703096].
 
\bibitem{Ali}See, for example, the review article and the references therein.\\
A. Mostafazadeh, {\em Pseudo-Hermitian representation of quantum mechanics}, Int. J. Geom. Meth. Mod. Phys. {\bf 7} (2010) 1191 [arXiv:0810.5643 [quant-ph]].

\bibitem{Swanson}
M.S. Swanson, {\em Transition elements for a non-Hermitian quadratic Hamiltonian}, J. Math. Phys. {\bf 45} (2004) 585;\\
C.M. Bender and H.F. Jones, {\em Interactions of Hermitian and non-Hermitian Hamiltonians},
J. Phys. {\bf A 41} (2008) 244006 [arXiv:0709.3605 [hep-th]].

\bibitem{PU}
A. Pais and G.E. Uhlenbeck, {\em On field theories with non-localized action}, Phys. Rev. {\bf 79} (1950) 145.

\bibitem{Bender1}C.M. Bender and P.D. Mannheim, {\em No-ghost theorem for the fourth-order derivative Pais-Uhlenbeck oscillator model},
Phys. Rev. Lett. {\bf 100} (2008) 110402 [arXiv:0706.0207 [hep-th]].

\bibitem{LM}
J.-Q. Li and  Y.-G. Miao, {\em Spontaneous breaking of permutation symmetry in pseudo-Hermitian quantum mechanics}, Phys. Rev. 
{\bf A 85} (2012) 042110 [arXiv:1110.2312 [quant-ph]].


\bibitem{Makris}
K.G. Makris, R. El-Ganainy, D.N. Christodoulides, and Z.H. Musslimani, {\em Beam dynamics in $\mathcal{PT}$ symmetric optical lattices}, Phys. Rev. Lett. {\bf 100} (2008) 103904.


\end{thebibliography}
\end{document}